\documentclass[epj]{svjour}
\usepackage{graphics}
\usepackage{youngtab}
\newcommand{\bea}{\begin{eqnarray}}
\newcommand{\eea}{\end{eqnarray}}
\newcommand{\nn}{\nonumber}
\begin{document}
\title{Excited baryons and heavy pentaquarks in large $N_c$ QCD}
\author{Dan Pirjol\inst{1} \and Carlos Schat\inst{2}% etc
\thanks{\emph{Present address:} Departamento de F\'{\i}sica, Universidad de Murcia, E30071 Murcia, Spain.}%
}                     % Do not remove
\institute{Center for Theoretical Physics, Massachusetts Institute of
Technology, Cambridge, MA 02139. \and  CONICET 
and Departamento de F\'{\i}sica, FCEyN, Universidad de Buenos Aires, \\
Ciudad Universitaria, Pab.1, (1428) Buenos Aires, Argentina.}
\date{Received: date / Revised version: date}
\abstract{We briefly discuss the large $N_c$ picture for excited baryons, 
present a new method for the calculation of matrix elements and  illustrate it 
by computing the
strong decays of heavy exotic states.
 % end of PACS codes
} %end of abstract
\maketitle
\section{Introduction}
\label{intro}
The $1/N_c$ expansion of QCD has turned out to be a fruitful
approach to its non-perturbative regime, as is shown by many examples
\cite{Goity:2005fj}.
The successful applications to the study of
ground state baryons make the excited baryons and exotic states especially 
interesting because they provide a wider testing ground for the $1/N_c$
expansion.

It is useful to
recall a few general facts that make the large number of colors limit
interesting and useful:
\begin{itemize}
%\item Although in the large $N_c$ limit the degrees of freedom increase, the physics simplifies.

\item The $1/N_c$ expansion is the only candidate for a perturbative expansion of QCD at all energies.

\item In the $N_c \rightarrow \infty$ limit baryons fall into
 irreducible representations of the {\it contracted} spin-flavor algebra $SU(2 n_f)_c$ ,
also known as ${\cal K}$-symmetry, that relates properties of states in different
multiplets of flavor symmetry.

\item The breaking of spin-flavor symmetry can be studied order by
order in $1/N_c$ as an operator expansion.

\end{itemize}

It is important to 
stress that already at leading order in the large $N_c$ limit it is possible to obtain 
significant insights into the structure of excited baryons, among which we would like to highlight the 
following:

\begin{itemize}

\item  The three towers \cite{Pirjol:1997sr} \cite{Pirjol:2003ye} \cite{Cohen:2003tb} predicted by ${\cal K}$-symmetry 
for the $L=1$ negative parity $N^*$ baryons , labeled by ${\cal K}=0,1,2$ with ${\cal K}$
related to the isospin $I$ and spin $J$ of the $N^*$'s by ${\cal K}\ge|I-J|$. 

\item The vanishing of the strong decay width  $\Gamma(N^*_{\frac12} \to [N \pi]_S)$
for $N^*_{\frac12}$ in the ${\cal K}=0$ tower, which provides a 
natural explanation for the relative suppresion of pion decays for the  
$N^*(1535)$  \cite{Pirjol:1997sr} \cite{Cohen:2003tb} \cite{Pirjol:2003qk}.

\item The order ${\cal O}(N_c^0)$ mass splitting of the $SU(3)$ singlets   $\Lambda(1405)$ - $\Lambda(1520)$
in the $[\mathbf{70},1^-]$ multiplet \cite{Schat:2001xr}. 

\end{itemize}

The general framework is based on the observation  that 
at the fundamental level of QCD diagrams can be classified according to their
scaling with $N_c$.  Planar diagrams are the leading order,
non-planar diagrams and quark loops are subleading in $1/N_c$. In
order to obtain finite amplitudes the quark-gluon coupling
constant must scale as $g\propto N_c^{-1/2}$.
An $m$ -body operator requires at least the exchange of $m-1$
gluons which gives a suppression factor of  $
N^{1-m}_c $. However, the matrix elements of an operator can eventually be
enhanced by coherence effects, as is the case of $G^{ia}$ defined below\footnote
{$\langle G^{ia} \rangle \propto N_c $ when restricted to the subspace of states with spin and isospin of order $N_c^0$, 
which are the ones that will correspond to the $N_c=3$ physical states}. 
In an explicit quark operator representation different hadronic operators like the masses,
magnetic moments, axial currents, etc.,   can be
expanded \cite{Dashen:1993jt} in $1/N_c$. For example, for the mass operator we have schematically
  \bea\label{massop} \hat M = \sum_{k=0}^{N_c} \frac{1}{N_c^{k-1}}
C_k {\mathcal O}_k \eea with ${\mathcal O}_k$ a $k$-body operator.
Both the coefficients $C_k$ (which correspond to  reduced matrix elements of QCD operators)
and the matrix elements of the
quark operators on baryon states $\langle {\mathcal O}_k\rangle$ have
power expansions in $1/N_c$ with coefficients determined by
nonperturbative dynamics.
The basic building blocks to construct the ${\mathcal O}_k$ are the generators of $SU(2 n_f)$, where
$n_f$ is the number of flavors
\begin{equation}
{ S^i }= \sum_{\alpha=1}^{N_c} s^i_{(\alpha)}, \; \;
  { T^a} = \sum_{\alpha=1}^{N_c} t^a_{(\alpha)}, \; \;
  { G^{ia}} = \sum_{\alpha=1}^{N_c} s^i_{(\alpha)}
  t^a_{(\alpha)} \ .
\label{bb}
\end{equation}
In the large $N_c$ limit we can define
$
X_{ia}^0 \equiv \lim_{N_c \rightarrow \infty} \frac{G_{ia}}{N_c}
$ ,
  because the matrix elements of ${G_{ia}}$
scale like $N_c$ for the states of interest, which is the
coherence effect mentioned before. In this way we obtain for $n_f=2$ the
contracted algebra $SU(4)_c$
\bea & & \left[S_i,S_j\right] =
i\epsilon_{ijk} S_k \ \ , \ \ \left[S_i,X_{ja}^0\right]  =
i\epsilon_{ijk} X_{ka}^0 \ ,
\nonumber \\
& & \left[T_a,T_b\right] = i \epsilon_{abc} T_c \ \ , \ \  \left[T_a,X_{ib}^0\right] =  i\epsilon _{abc} X_{ic}^0 \ ,
\nonumber \\
& & \left[X_{ia}^0,X_{jb}^0\right]  =  0 \ .
\label{xx}
\eea
The last commutation relations can also be obtained in a purely hadronic language.
They are known as consistency relations \cite{Dashen:1993jt} and are necessary to obtain finite
amplitudes for pion-nucleon scattering. Consider the direct and crossed diagrams that contribute at tree level.
The pion-nucleon coupling scales like $\sqrt{N_c}$, which makes each diagram separately to scale
like $N_c$. To obtain a finite amplitude for the physical process we need a cancellation to happen.
This requires $\left[X_{ia}^0,X_{jb}^0\right]  =  {\mathcal O}(1/N_c)$, which in the large $N_c$ limit
gives Eq.(\ref{xx}). This symmetry structure gives rise to model independent predictions like the three
towers for excited baryons that was mentioned above. In an explicit quark operator representation this 
is manifest by the presence of two ${\cal O}(N_c^0)$ operators (that also involve 
the generator of $O(3)$ \cite{Goity:1996hk}) and has been checked by an explicit calculation 
\cite{Pirjol:2003ye} \cite{Cohen:2003tb}. 

\section{Occupation number formalism}
\label{sec:11}

In this section we give an outline of the occupation number formalism \cite{Pirjol:2006jp} that 
we use to compute matrix elements for arbitrary $N_c$. 
In broken $SU(3)$, the $SU(6)$ spin-flavor  generators
can be decomposed into generators of the subgroup
\begin{eqnarray}\label{decomp}
SU(6)_{SF} \supset SU(4)_{SI} \otimes SU(2)_{J_s} \otimes U(1)_{n_s} \ \nn 
\end{eqnarray}
\begin{eqnarray}\label{bblocks}
&& J^i  \ \ , \ \ I^a = T^a \ \ , \ \ G^{ia} = G^{ia}  \qquad (i,a = 1...3)  \ , 
\nonumber \\
&& J_s^i = s^\dagger \frac{\sigma^i}{2}  s \ , \ N_s = s^\dagger s \, \nonumber
\end{eqnarray}
plus operators mediating transitions between sectors of different
$n_s$
\begin{eqnarray}\label{bblocks2}
&& \tilde t^\alpha =  q^{\dagger\alpha} s \ \ , \ \
t_\alpha = s^\dagger q_\alpha \, \ \ \ \ \ \ \ \ \ \  \quad (\alpha = \pm 1/2) \ , \nonumber \\
&& \tilde Y^{i\alpha} =  q^{\dagger\alpha} \frac{\sigma^i}{2} s
\ \ , \ \ 
Y^i_\alpha = s^\dagger \frac{\sigma^i}{2} q_\alpha  \ . \nn 
\end{eqnarray}

\noindent
We introduce the ``6n-symbol'' defined as ($N = \sum_{i=1}^6 n_i$)
\begin{eqnarray}\label{quark}
&& \{ n_1, n_2, n_3, n_4, n_5, n_6 \} = 
\sqrt{\frac{n_1 ! n_2 ! n_3 ! n_4 ! n_5! n_6!}{N !}} \nn \\
&& \qquad \qquad \times
 (u_\uparrow^{n_1}u_\downarrow^{n_2}d_\uparrow^{n_3}d_\downarrow^{n_4}
s_\uparrow^{n_5}s_\downarrow^{n_6} + \mbox{perms})\, . \nn
\end{eqnarray}

The nonstrange states in a ${\cal K}=0$ tower have spin and
isopin satisfying $I=J$. Their spin-flavor symmetric wave functions can be given in closed form
as 
\begin{eqnarray}\label{nons}
&& |I I_3 J_3 ; N_{ud} \rangle = \sum_i
\left(
\begin{array}{cc|c}
\frac{N_u}{2}& \frac{N_d}{2} & I \\
i & J_3-i & J_3 \\
\end{array}
\right) \nn \\
&& \qquad \qquad \times
\{ \frac{N_u}{2}+i, \frac{N_u}{2}-i,
\frac{N_d}{2}+J_3-i,
\frac{N_d}{2}-J_3+i\} \ , \nn
\end{eqnarray}
where $N_{u,d}$ are the number of up and down quarks:
$
N_u = \frac{N_{ud}}{2}+I_3\,, 
N_d = \frac{N_{ud}}{2}-I_3 \ 
$
with $N_{ud}=N_c-n_s$. \\
A few representative nonstrange $J_3 = +\frac12$ states are
\begin{eqnarray}
p_\uparrow = \sqrt{\frac23} \{2,0,0,1\} - \frac{1}{\sqrt3}
\{1,1,1,0\} \ &,& 
\Delta^{++}_\uparrow = \{2, 1, 0, 0\} \ . \nn
\end{eqnarray}
\noindent
Acting with 
\begin{eqnarray}\label{rules}
&& q_i \{ \cdots, n_i, \cdots \} =
\sqrt{n_i} \{ \cdots, n_i-1, \cdots \} \ , \nn \\ 
&& q_i^\dagger \{ \cdots, n_i, \cdots \} =
\sqrt{n_i+1} \{ \cdots, n_i+1, \cdots \} 
\end{eqnarray}
we obtain the matrix elements of any operator for arbitrary $N_c$.

\section{Pentaquark towers}
\label{sec:19}

For the exotic $q^{N_c+1} \bar q$ states with $N_c+1$ quarks in a $\mathbf{``\bar 3"}$ of 
color, Fermi statistics implies the 
$SU(6)\otimes O(3)$ decomposition 
{\bf 
\begin{eqnarray}\label{spflorb}
&& \raisebox{-0.4cm}{\yng(5,1)}\quad  \to  \nn \\
&&
\left[ \quad  \yng(6)\quad \otimes\quad
\raisebox{-0.4cm}{\yng(5,1)} \quad \right]_{\rm parity +} \oplus  \nn \\
& & 
\left[\quad  \raisebox{-0.4cm}{\yng(5,1)}
\quad \otimes\quad \yng(6)\quad \right]_{\rm parity -} 
\oplus  \cdots\, \nonumber
\end{eqnarray}
}
The negative parity states were studied in \cite{Pirjol:2004dw}. Here we reconsider the 
positive parity states \cite{Jenkins:2004vb}, which are all members of the two towers

\vspace{.1cm}
$
\begin{array}{ccc}\label{K12}
{\cal K} =1/2: && \mathbf{\overline{10}}_\frac12\,,\quad
\mathbf{27}_{\frac12, \frac32}\,,\quad \quad \mathbf{35}_{\frac32, \frac52}\,,\cdots
\nn \\
\label{K32}
{\cal K} =3/2: &&  \quad \ \mathbf{\overline{10}}_\frac32\,,\quad
\mathbf{27}_{\frac12, \frac32, \frac52}\,,\quad 
\mathbf{35}_{\frac12, \frac32, \frac52, \frac72}\,,\cdots\, \nn
\end{array}
$
\vspace{.1cm}

In \cite{Jenkins:2004vb} only  states in the first tower were considered.
In the heavy quark limit $m_Q \rightarrow \infty$ these two towers become degenerate 
and the tower label for the light degrees of freedom becomes a good quantum number
\begin{eqnarray}\label{set1}
{\cal K}_{light}=1: && \mathbf{\overline{6}}_1\,,\quad
\mathbf{15}_{0,1,2}\,,\quad \mathbf{15'}_{1,2,3}\, \quad, \quad \cdots 
\end{eqnarray}
On the other hand the heavy pentaquarks considered in \cite{Jenkins:2004vb} belong to the tower 
\begin{eqnarray} \label{set0}
{\cal K}_{light}=0: &&\mathbf{\overline{6}}_0\,, \quad \mathbf{15}_{1}\,,\quad  \quad \
\mathbf{15'}_{2}\, \quad, \quad \cdots \
\end{eqnarray}
which arises naturally in the Skyrme model. 

\begin{table}
\caption{Reduced matrix elements $Y$
and large $N_c$ width for the ${\cal K}=1/2$ pentaquark $\Theta_{{\bar Q} J_\ell} \to NK, \Delta K$ decays.
} 
\label{tab:1} 
%\begin{ruledtabular}
\begin{tabular}{lllc}
\hline\noalign{\smallskip}
Decay & $(I'J',IJ)$ & $Y(I'J'{\cal K'},IJ{\cal K})$ & $\frac{1}{p^3} {\rm \Gamma}_{N_c \to \infty}^{\rm p-wave}$ \\
\noalign{\smallskip}\hline\noalign{\smallskip}
$\Theta_0(\frac12) \to NK$ & $( \frac12\frac12, 0\frac12)$  &
                         $-\frac{\sqrt3}{2}\sqrt{N_c+1}$ & $1$ \\
%\hline
$\Theta_1(\frac12) \to NK$ & $(\frac12\frac12, 1\frac12)$  &
                             $\frac12\sqrt{N_c+5}$ &
                             $\frac19$ \\
$\qquad \quad \to \Delta K$ & $(\frac32\frac32, 1\frac12)$  &
                             $\frac{1}{\sqrt2}\sqrt{N_c-1}$ & $\frac89$ \\
%\hline
$\Theta_1(\frac32) \to NK$ & $(\frac12\frac12, 1\frac32)$  &
                             $-\sqrt2\sqrt{N_c+5}$ & $\frac49$ \\
$ \qquad \quad\to \Delta K$ & $(\frac32\frac32, 1\frac32)$  &
                 $-\frac12\sqrt{\frac52}\sqrt{N_c-1}$ &
                 $\frac59$ \\
%\hline
\noalign{\smallskip}\hline
\end{tabular}
%\end{ruledtabular}
\end{table}
As an example we compute the strong decays of the ${\cal K}=1/2$ states in \cite{Jenkins:2004vb}.
The reduced matrix elements of the 
transition operator are defined by 
\begin{eqnarray}\label{Ttildedef}
&& \langle I' I_3' , J'J_3' ;n_s-1| Y^{i\alpha }| I I_3 , J J_3 ;n_s\rangle
=\\
&& \qquad \qquad
\left(
\begin{array}{cc|c}
I &  \frac12 &  I'\\
I_3 &  \alpha & I'_3
\end{array}
\right)
\left(
\begin{array}{cc|c}
J &  1 &  J'\\
J_3 & i & J'_3
\end{array}
\right)
Y(I'J'{\cal K'},IJ{\cal K})
\nonumber
\end{eqnarray}
In the large $N_c$ limit we find \cite{PirSch}
\begin{eqnarray}\label{Ksol}
&& Y_0(I'J'{\cal K'},IJ{\cal K}) 
\propto \sqrt{[I][J]}
\left\{
\begin{array}{ccc}
\frac12 & 1 & \frac12 \\
I & J & {\cal K} \\
I' & J' & {\cal K'} \\
\end{array}
\right\} \ .
\end{eqnarray}
The expressions for arbitrary $N_c$ are found in Table~\ref{tab:1}.
Averaging over initial states and summing over final states the p-wave widths are obtained as
\begin{eqnarray}
{\rm \Gamma}(I'J'{\cal K'},IJ{\cal K}) &\propto &  \frac{[I'][J']}{[I][J]}
|{Y}(I'J'{\cal K'},IJ{\cal K})|^2 \ . \nn
\end{eqnarray}
In the large $N_c$ limit all pentaquark states in the same tower have the same
total width.  This leads to sum rules like
{\small 
\begin{eqnarray}\nn
{\rm \Gamma}(\Theta_0(\frac12) \to NK)&=&{\rm \Gamma}(\Theta_1(\frac12) \to NK) +
{\rm \Gamma}(\Theta_1(\frac12) \to \Delta K) \\ \nn
&=&{\rm \Gamma}(\Theta_1(\frac32) \to NK) + {\rm \Gamma}(\Theta_1(\frac32) \to \Delta K) 
\end{eqnarray}
}
as can be verified from Table~\ref{tab:1}. The results for $N_c=3$ in \cite{Jenkins:2004vb}
can also be  verified from Table~\ref{tab:1}.

\section{ Large $N_c$ and heavy quark limit predictions}
\label{sec:14}

\begin{table}
\caption{Heavy quark symmetry predictions 
for the decay amplitudes  
$\Theta_{{\bar Q} J_\ell} \to [N H_{\bar Q}^{(*)}]_{\rm p-wave}$. }
\label{tab:2}       % Give a unique label
\begin{tabular}{lcc}
\hline\noalign{\smallskip}
Decay & $J_N=1/2$ & $J_N = 3/2$ \\
\noalign{\smallskip}\hline\noalign{\smallskip}
$\Theta_{{\bar Q}0}(\frac12) \to NH_{\bar Q}$ & $-\frac12 f_0$ & $-$ \\
$\Theta_{{\bar Q}1}(\frac12) \to NH_{\bar Q}$ & $\frac{\sqrt3}{2} f_1$ & $-$ \\
$\Theta_{{\bar Q}1}(\frac32) \to NH_{\bar Q}$ & $-$ & $-\frac12 \sqrt{\frac32} f_2$ \\
\noalign{\smallskip}\hline\noalign{\smallskip}
$\Theta_{{\bar Q}0}(\frac12)\to NH^*_{\bar Q}$ & $\frac{\sqrt3}{2}f_0$ & $-$ \\
$\Theta_{{\bar Q}1}(\frac12)\to NH^*_{\bar Q}$ & $\frac{1}{2}f_1$ & $-f_2$ \\
$\Theta_{{\bar Q}1}(\frac32)\to NH^*_{\bar Q}$ & $-f_1$ & $\frac12\sqrt{\frac52}f_2$ \\
\noalign{\smallskip}\hline
\end{tabular}
% Or use
%%%\vspace*{5cm}  % with the correct table height
\end{table}

Heavy quark symmetry predicts the amplitudes in terms of a few reduced 
matrix elements {$f_i$}. The decay amplitude for 
$\Theta_{\bar Q}(IJ J_\ell) \to [N H^{(*)}_{\bar Q}(J'J'_\ell)]_{J_N}$, 
where ${\bf J}_N = {\bf S}_N + {\bf L}$ is the angular momentum carried 
by the final baryon, is given by \cite{Isgur:1991wq}
\begin{eqnarray}
A_i = \sqrt{(2J_\ell +1)(2J'+1)}
\left\{
\begin{array}{ccc}
J_\ell & J'_\ell & J_N \\
J' & J & \frac12 \\
\end{array}
\right\} f_i \ .
\end{eqnarray}
Combining the heavy quark symmetry predictions with the large $N_c$ amplitudes
we can fix the reduced 
amplitudes {$f_i$} and obtain predictions for the ratios of decays widths, 
as summarized for the $I=1$ states in Table~\ref{tab:3}. More details will be given elsewhere \cite{PirSch}.

\begin{table}
\caption{Ratios of strong decay widths for heavy pentaquarks $R^I(J)= \Theta_{\bar Q}^I (J)\to (N H_{\bar Q}) : (N H_{\bar Q}^*)$ }
\label{tab:3}       % Give a unique label
\begin{tabular}{lll}
\hline\noalign{\smallskip}
& & \\[-0.12in]
 {$I=1$} & $R^I(J=\frac12)$ 
  & $R^I(J=\frac32)$  \\[0.08in]
% & $\Gamma(\Theta_Q^{(I=0)}(\frac12) \to N\bar D) : (N\bar D^*)$ & 
%   $\Gamma(\Theta_c^{(I=1)}(\frac12) \to N\bar D) : (N\bar D^*)$ & 
%   $\Gamma(\Theta_c^{(I=1)}(\frac32) \to N\bar D) : (N\bar D^*)$ \\
\noalign{\smallskip}\hline\noalign{\smallskip}
& & \\[-0.12in]
${\cal K}_{light}=1$ \quad \quad  &  $1$ : 3 \hspace{0.1cm} $(J_\ell = 0)$ \quad  
          &  $\frac12$ : $\frac72$ \hspace{0.1cm} $(J_\ell = 1)$     \\[0.08in]
 &  $2$ : $2$ \hspace{0.1cm} $(J_\ell = 1)$  
          &  $\frac52$ : $\frac32$ \hspace{0.1cm} $(J_\ell = 2)$     \\[0.08in]
\hline
& & \\[-0.12in]
${\cal K}_{light}=0 $    
           & $1$ : $11$ \hspace{0.1cm} $(J_\ell = 1)$ 
           & $4$ : $8$ \hspace{0.1cm} $(J_\ell = 1)$ \\[0.08in]
\noalign{\smallskip}\hline
\end{tabular}
% Or use
%%%\vspace*{5cm}  % with the correct table height
\end{table}

\section{Conclusions}
\label{sec:conclusions}

The large $N_c$ limit reveals a structure of mass degeneracies 
and sum rules for decay widths that is not apparent at $N_c=3$.
This picture can be corrected systematically
in 1/$N_c$. We presented a new method for computing matrix elements 
for arbitrary $N_c$ which is useful for this purpose. As an illustration, we 
showed how the combined large $N_c$ and heavy quark limit allows 
to compute decay width ratios that discriminate between different heavy pentaquark 
states. In the heavy quark limit the spin of the light degrees of freedom is a conserved 
quantum number. When this is combined with the large $N_c$ limit we can label the
states by the new quantum number ${\cal K}_{light}$. The states considered in \cite{Jenkins:2004vb}
have ${\cal K}_{light}=0$ while the states considered in this work have ${\cal K}_{light}=1$. 
The predictions for their strong decays differ, as can be seen in Table~\ref{tab:3}. \\

{\em Acknowledgements:} The work of C.S. was supported in part by Fundaci\'on Antorchas, Argentina 
and Fundaci\'on S\'eneca, Murcia, Spain. C.S. thanks for the hospitality of the Departamento de F\'{\i}sica, 
Universidad de Murcia during the completion of this work.

%
% BibTeX users please use
% \bibliographystyle{}
% \bibliography{}

\end{document}